\documentclass[useAMS,usenatbib]{mn2e}
\usepackage[dvips]{graphicx}
\usepackage{psfig}
\usepackage{lscape}
\usepackage{supertabular}

\title[New nearby L, and late-M dwarfs]{Discovery of new nearby L and late-M dwarfs at low Galactic latitude from the DENIS database.}
\author[N. Phan-Bao et al.]{N. Phan-Bao$^{1,2}$\thanks{E-mail: pbngoc@asiaa.sinica.edu.tw(PBN)}, 
  M.S. Bessell$^{3}$,
  E.L. Mart\'{\i}n$^{4,2}$,
  G. Simon$^{5}$,
  J. Borsenberger$^{6}$,
      \newauthor
  R. Tata$^{2}$ 
  J. Guibert$^{7}$,
  F. Crifo$^{5}$,
  T. Forveille$^{8}$,
  X. Delfosse$^{8}$,
  J. Lim$^{1}$,
  B. de Batz$^{5}$ \\
$^{1}$Institute of Astronomy and Astrophysics, Academia Sinica. 
      PO Box 23-141, Taipei 106, Taiwan, ROC. \\
$^{2}$University of Central Florida, Dept. of Physics, PO Box 162385, 
          Orlando, FL 32816-2385, USA. \\
$^{3}$Research School of Astronomy and Astrophysics, Australian National 
           University, Cotter Rd, Weston, ACT 2611, Australia. \\
$^{4}$Instituto de Astrof\'{\i}sica de Canarias, C/ V\'{\i}a L\'actea  
      s/n, E-38200 La Laguna (Tenerife), Spain. \\
$^{5}$GEPI, Observatoire de Paris, 5 place J. Janssen, 92195 
      Meudon Cedex, France. \\
$^{6}$LESIA, Observatoire de Paris, 5 place J. Janssen, 92195 
      Meudon Cedex, France. \\
$^{7}$Centre d'Analyse des Images, GEPI, Observatoire de Paris,  
      61 avenue de l'Observatoire, 75014 Paris, France. \\
$^{8}$Laboratoire d'Astrophysique de Grenoble, Universit\'e J. 
      Fourier, B.P. 53, F-38041 Grenoble, France. \\
}

\begin{document}
      \date{Received / Accepted}
\pagerange{\pageref{firstpage}--\pageref{lastpage}} \pubyear{2007}      

\maketitle

\label{firstpage}

\begin{abstract}

We report new nearby L and late-M dwarfs
($d_{\rm{phot}}~\leq~30$~pc) discovered in our search for
nearby ultracool dwarfs ($I-J \geq 3.0$, later than 
M8.0) at low Galactic latitude ($|b| < 15\degr$) 
over 4,800 square degrees in the DENIS database.
We used late-M ($\geq$M8.0), 
L, and T dwarfs with accurate trigonometric parallaxes 
to calibrate the $M_{\rm J}$ versus $I-J$ colour-luminosity 
relation. The resulting photometric distances have standard errors of 
$\sim 15\%$, which we used to select candidates $d_{\rm{phot}}~\leq~30$~pc.
We measured proper motions from
multi-epoch images found in the public archives ALADIN, DSS, 
2MASS, DENIS, with at least three distinct epochs and 
time baselines of 10 to 21~years. We then used a Maximum Reduced Proper 
Motion cutoff to select 28 candidates as ultracool dwarfs 
(M8.0--L8.0) and to reject one as a distant red star. 
No T dwarf candidates were found in this search which required an object to be 
detected in all three DENIS bands.
Our low-resolution optical spectra confirmed that 26 of them were indeed ultracool
dwarfs, with spectral types from M8.0 to L5.5. Two contaminants and one rejected
by the Maximum Reduced Proper Motion cutoff were all reddened F-K main sequence 
stars. 20 of these 26 ultracool dwarfs are new nearby ultracool dwarf members, 
three L dwarfs within 15~pc with one L3.5 at only $\sim$10~pc.  
We determine a stellar density of $\overline{\Phi}_{\rm J~cor}=(1.64 \pm 0.46).10^{-3}$ dwarfs pc$^{-3}$ mag$^{-1}$
over $11.1 \leq M_{\rm J} \leq 13.1$ 
based on that sample of M8--L3.5 ultracool dwarfs. Our ultracool dwarf density value 
is in good agreement with the Cruz et al. measurement of 
the ultracool dwarf density at high Galactic latitude.

\end{abstract}


\section{INTRODUCTION}

Nearby stars are the brightest representatives of their 
class, and therefore provide observational benchmarks
for stellar physics. This is particularly true for 
intrinsically faint objects, such as stars
at the bottom of the main sequence, and brown dwarfs.

In the last decade, many nearby ultracool dwarfs have been discovered by using the DENIS \citep{epchtein97}, 2MASS \citep{skrutskie},  
SDSS \citep{york}, UKIDSS \citep{lawrence} surveys
(\citealt{delfosse99,martin99,p01,p03,reyle}; \citealt{kirk99, reid02, cruz03, bur04};
\citealt*{deacon05}; \citealt{lodieu05, lodieu07}; \citealt{knapp04, chiu})
or the proper motion measurement (\citealt*{lepine02}).

However, most surveys for nearby ultracool dwarfs have tended to avoid 
the crowded regions of the Galactic plane.
The first systematic search for ultracool dwarfs
($|b| < 10\degr$) was carried out by \citet{reid03}, 
he discovered one M8 and reidentified one L1.5 previously discovered by \citet{salim}.
\citet{ham04} reported an M9.0 dwarf 1RXS~J115928.5$-$524717 ($b = 9.3\degr$)
showing strong X-ray flaring emission. Recently, \citet{folkes} discovered
an L-T transition object close in the Galactic plane 2MASS~J11263991$-$5003550 
(L9, $b = 10.6\degr$). Four other ultracool dwarfs at low Galactic latitudes ($|b| < 15\degr$)
were also discovered: one M8 (\citealt{cruz03}); one L2 (\citealt{scholz02});
one L0 and one L4.5 (\citealt{kendall}), the latter L4.5 dwarf  
is actually an L1+L3.5 \citep{kendall} or L1.5+L4.5 \citep{burl} binary.

These discoveries of ultracool dwarfs in the Galactic plane provide
additional targets with which to study the basic physical properties of stars 
at the bottom of the main sequence and brown dwarfs; however, many remain to be discovered.  

In our previous work on an ultracool dwarf search at high Galactic latitude 
($|b| \geq 15\degr$),
we \citep{p03} used the Maximum Reduced Proper 
Motion (MRPM) method to select mid-M dwarfs 
($2.0 \leq I-J \leq 3.0$, $\sim$M6.0--M8.0) in the DENIS
database. This allowed us to identify not only high proper
motion (high-pm, $\mu \geq 0.1\arcsec$yr$^{-1}$) M dwarfs, but also some with low 
proper motion (low-pm, $\mu < 0.1\arcsec$yr$^{-1}$), such as DENIS-P~J1538317$-$103850 
(M5.0, $\mu=20$~mas~yr$^{-1}$). Here we apply the same 
technique to a search for nearby dwarfs cooler than 
M8.0 ($I-J \geq 3.0$) at low Galactic latitude ($|b| < 15\degr$)
in the whole DENIS database (90$\%$ complete of the DENIS survey).
This corresponds to  
a sky surface coverage of $\sim$4,800 square degrees 
or 44.9$\%$ of the Galactic plane ($|b| < 15\degr$). 
In this search, we discovered 20 new ultracool dwarfs and recovered 6 previously known ones,
giving a significant fraction of 77$\%$ of new ultracool dwarfs in our sample.

We present our $M_{\rm J}$ versus $I-J$ colour-magnitude relation in 
Section~2, and review the ultracool dwarf candidate selection in 
Section~3. We describe
the proper motion measurements in Section~4, and the MRPM filtering
of the candidates in Section~5. 
We present our low-resolution optical spectra and further spectral analysis in Section~6. 
We discuss the completeness of the sample in Section~7 and give a summary in Section~8.
\begin{figure}
\psfig{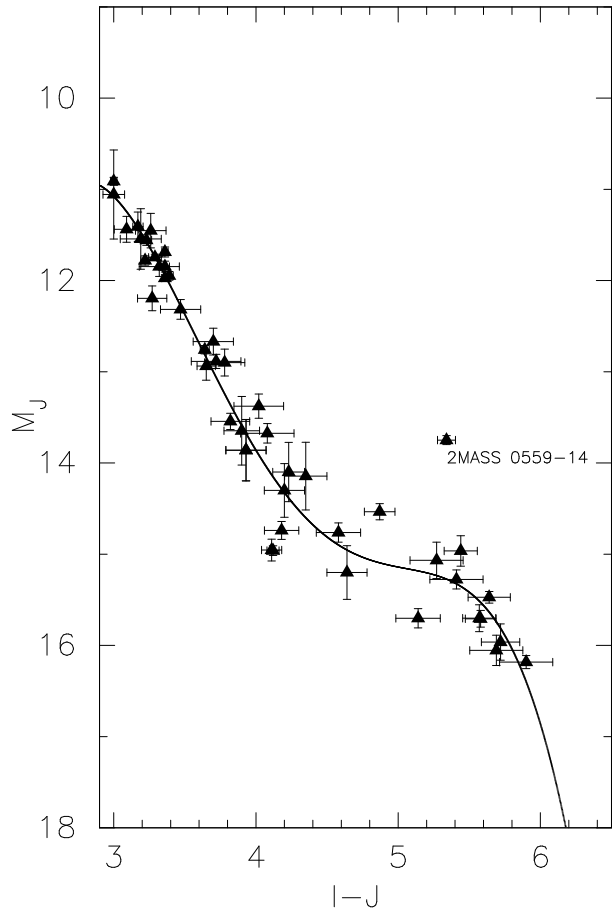}
\caption{($M_{\rm J}$, $I-J$) colour-magnitude diagram for late-M 
($\geq$M8.0) and L dwarfs with known trigonometric
parallaxes (data in Table~\ref{cali}). 2MASS 0559$-$1404 (T5.0, 
\citealt{bur02}) is overluminous, by more than the 0.75~magnitude
 offset expected for an equal mass binary system.
\label{MJ}}
\end{figure}
\begin{figure}
\psfig{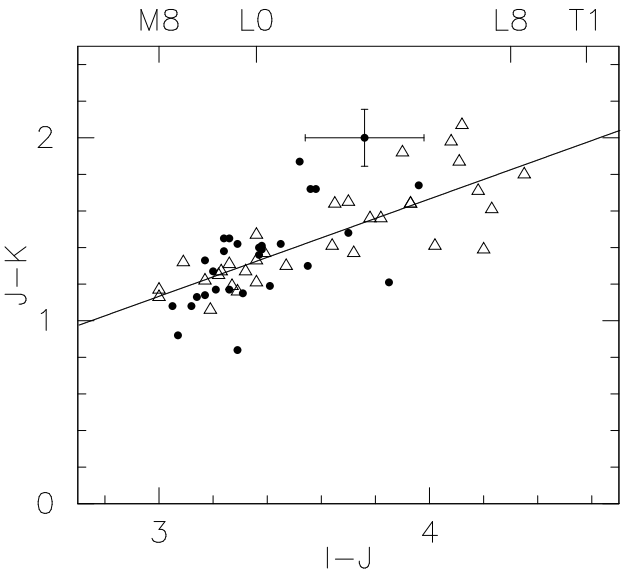}
\caption{($I-J$, $J-K$) colour-colour diagram for our 29 late-M 
and L dwarf candidates (Table~\ref{candidates}), plotted as 
solid circles, as well as known late-M and L dwarfs 
(Table~\ref{cali}); plotted as open triangles). Representative errorbars for one
object are plotted.
The line represents our linear fit to the colours of 
the literature dwarfs: $J-K = -0.462 + 0.532(I-J), \sigma=0.15$.  
\label{IJ_JK}}
\end{figure}
\section{COLOUR -- MAGNITUDE RELATION}
We identified 47 very cool and nearby ($d < 32$ pc) dwarfs in the literature with 
trigonometric parallaxes and good photometry (e.g. errors smaller than 0.2~mag)
(Table~\ref{cali}). For our limited present purpose 
the DENIS system is close enough to both the Cousins-CIT 
($\leq 0.05$~mag, \citealt{del97b}) and 2MASS systems 
($\leq 0.02$~mag, \citealt{car}). We therefore ignored 
the small corrections needed to transform between these 
photometric systems. We did, on the other hand, 
transform the $I_{814}$ magnitudes of \citep{reid01,bur03} 
to $I_{\rm C}$, using the linear correction of \citet{reid01}. 

Figure~\ref{MJ} shows the resulting ($I-J$, $M_{\rm J}$) plot, 
and the corresponding $4^{th}$ order polynomial fit:
\begin{eqnarray}
 M_{\rm J} & = & a_{0}+a_{1}(I-J)+a_{2}(I-J)^{2}+a_{3}(I-J)^{3} \nonumber \\ 
           &   & +~a_{4}(I-J)^{4} \label{eq1}
\end{eqnarray}
where $a_{0}=130.164$, $a_{1}=-124.899$, 
$a_{2}=46.948$, $a_{3}=-7.4842$, $a_{4}=0.4341$, 
valid for $3.0~\leq~I-J~\leq~6.0$. The rms dispersion
around that fit is 0.30~mag, corresponding to a 14\% error 
on distances.

\begin{table*}
   \caption{47 late-M, L, and T dwarfs with known trigonometric parallaxes}
   \label{cali}
  $$
 \begin{tabular}{llllllrlllrrl}
   \hline 
   \hline
   \noalign{\smallskip}
Stars  & $\alpha_{\rm 2000}$ & $\delta_{\rm 2000}$ & SpT & $I$ & $I-J$ & $J-K$ & err$I$ & err$J$ & err$K$ & pi & errpi & References \\ 
       &   &   &   &   &   &   &   &   &   & mas & mas &  \\ 
 (1)   & (2) & (3) & (4) & (5) & (6) & (7) & (8) & (9) & (10) & (11) & (12) & (13) \\
   \hline
   \noalign{\smallskip}  
LHS 102B           & 00 04 34.68 & $-$40 44 01.7 & L4.5 & 17.69 & 3.93 &   1.64 & 0.10 & 0.10 & 0.10 &  104.7 &  11.4  & 1,2,2,2,3     \\
LHS 102C           & 00 04 34.68 & $-$40 44 01.7 & L4.5 & 17.69 & 3.93 &   1.64 & 0.10 & 0.10 & 0.10 &  104.7 &  11.4  & 1,2,2,2,3     \\
BRI 0021$-$0214    & 00 24 24.63 & $-$01 58 20.0 & M9.5 & 15.13 & 3.26 &   1.31 & 0.04 & 0.10 & 0.09 &   82.5 &   3.4  & 4,2,2,2,3     \\
2M 0030$-$1450     & 00 30 30.13 & $-$14 50 33.4 & L7.0 & 20.63 & 4.35 &   1.80 & 0.10 & 0.11 & 0.10 &   37.4 &   4.5  & 5,6,7,7,8     \\
2M 0036$+$1821     & 00 36 16.18 & $+$18 21 10.5 & L3.5 & 16.11 & 3.64 &   1.41 & 0.01 & 0.03 & 0.02 &  114.2 &   0.8  & 9,10,7,7,10   \\
RGO 0050$-$2722    & 00 52 54.67 & $-$27 05 59.5 & M9.0 & 16.67 & 3.19 &   1.06 & 0.08 & 0.12 & 0.14 &   41.0 &   4.0  & 11,2,2,2,3    \\
2M 0149$+$29       & 01 49 08.96 & $+$29 56 13.2 & M9.5 & 16.81 & 3.36 &   1.47 & 0.01 & 0.02 & 0.02 &   44.4 &   0.7  & 12,10,7,7,10  \\
T* 832-10443       & 02 52 26.29 & $+$00 56 22.4 & M8.0 & 16.13 & 3.00 &   1.17 & 0.01 & 0.02 & 0.02 &   36.0 &   0.4  & 13,10,7,7,10  \\
2M 0326$+$2950     & 03 26 13.67 & $+$29 50 15.2 & L3.5 & 19.13 & 3.65 &   1.64 & 0.04 & 0.05 & 0.05 &   31.0 &   1.5  & 14,10,7,7,10  \\
2M 0328$+$2302     & 03 28 42.66 & $+$23 02 05.2 & L8.0 & 20.73 & 4.23 &   1.61 & 0.10 & 0.05 & 0.05 &   33.1 &   4.2  & 5,6,15,15,8   \\
LP 944-20          & 03 39 35.26 & $-$35 25 43.6 & M9.0 & 13.96 & 3.27 &   1.19 & 0.05 & 0.09 & 0.10 &  200.0 &   4.2  & 4,2,2,2,16    \\
2M 0345$+$25       & 03 45 43.16 & $+$25 40 23.3 & L0.0 & 17.36 & 3.36 &   1.33 & 0.01 & 0.03 & 0.02 &   37.1 &   0.5  & 14,17,7,7,10  \\
S* 0539$-$0059     & 05 39 52.00 & $-$00 59 01.9 & L5.0 & 17.99 & 4.02 &   1.41 & 0.16 & 0.07 & 0.15 &   76.1 &   2.2  & 18,2,2,2,8    \\
2M 0559$-$1404     & 05 59 19.14 & $-$14 04 49.0 & T5.0 & 19.14 & 5.34 &   0.22 & 0.06 & 0.02 & 0.05 &   97.7 &   1.3  & 19,10,7,7,10  \\
Gl 229B            & 06 10 34.70 & $-$21 51 46.0 & T6.5 & 19.92 & 5.64 &$-$0.04 & 0.14 & 0.05 & 0.05 &  173.2 &   1.1  & 19,10,15,15,20 \\
2M 0746$+$2000A    & 07 46 42.56 & $+$20 00 32.2 & L0.5 & 15.60 & 3.32 &   1.24 & 0.10 & 0.10 & 0.07 &   81.9 &   0.3  & 21,22,22,21,10 \\
2M 0746$+$2000B    & 07 46 42.56 & $+$20 00 32.2 & L2.0 & 16.22 & 3.47 &   1.27 & 0.10 & 0.10 & 0.16 &   81.9 &   0.3  & 21,22,22,21,10 \\
2M 0825$+$2115     & 08 25 19.68 & $+$21 15 52.1 & L7.5 & 19.22 & 4.12 &   2.07 & 0.03 & 0.03 & 0.03 &   93.8 &   1.0  & 5,10,7,7,10   \\
2M 0850$+$1057A    & 08 50 35.93 & $+$10 57 15.6 & L6.0 & 20.54 & 4.20 &   1.39 & 0.10 & 0.10 & 0.10 &   39.1 &   3.5  & 22,22,22,7,10 \\
2M 0850$+$1057B    & 08 50 35.93 & $+$10 57 15.6 & L8.0 & 21.88 & 4.64 &   1.64 & 0.10 & 0.10 & 0.10 &   39.1 &   3.5  & 22,22,22,7,10 \\
LHS 2065           & 08 53 36.20 & $-$03 29 32.1 & M9.0 & 14.44 & 3.23 &   1.27 & 0.02 & 0.03 & 0.02 &  117.3 &   1.5  & 11,23,7,7,24  \\
2M 0937$+$2931     & 09 37 34.88 & $+$29 31 41.0 & T6.0 & 20.23 & 5.58 &$-$0.62 & 0.10 & 0.04 & 0.13 &  162.8 &   3.9  & 19,25,7,7,8   \\
2M 1047$+$2124     & 10 47 53.85 & $+$21 24 23.5 & T6.5 & 21.39 & 5.57 &$-$0.59 & 0.10 & 0.06 & 0.10 &   94.7 &   3.8  & 19,25,7,7,8   \\
D* 1048$-$3956     & 10 48 14.42 & $-$39 56 08.2 & M9.0 & 12.64 & 3.00 &   1.13 & 0.03 & 0.07 & 0.07 &  192.0 &  37.0  & 26,2,2,2,27   \\
D* 1058$-$1548     & 10 58 46.50 & $-$15 48 00.0 & L2.5 & 17.80 & 3.72 &   1.37 & 0.17 & 0.04 & 0.14 &   57.7 &   1.0  & 11,2,2,2,10   \\
2M 1146$+$2230A    & 11 46 34.49 & $+$22 30 52.7 & L3.0 & 18.54 & 3.70 &   1.65 & 0.10 & 0.10 & 0.10 &   36.8 &   0.8  & 28,22,21,7,8  \\
2M 1146$+$2230B    & 11 46 34.49 & $+$22 30 52.7 & L4.0 & 18.85 & 3.78 &   1.56 & 0.10 & 0.10 & 0.10 &   36.8 &   0.8  & 28,22,21,7,8  \\
2M 1217$-$0311     & 12 17 11.10 & $-$03 11 13.1 & T7.5 & 21.53 & 5.69 &   0.03 & 0.18 & 0.05 & 0.05 &  110.4 &   5.9  & 19,10,15,15,8 \\
BRI 1222$-$1221    & 12 24 52.23 & $-$12 38 35.2 & M8.5 & 15.74 & 3.17 &   1.22 & 0.03 & 0.02 & 0.03 &   58.6 &   3.8  & 11,23,7,7,16  \\
2M 1237$+$6526     & 12 37 39.20 & $+$65 26 14.8 & T6.5 & 21.77 & 5.72 &$-$0.01 & 0.10 & 0.09 & 0.20 &   96.1 &   4.8  & 19,25,7,7,8   \\
S* 1254$-$0122     & 12 54 53.90 & $-$01 22 47.4 & T2.0 & 19.76 & 4.87 &   1.05 & 0.10 & 0.04 & 0.05 &   84.9 &   1.9  & 19,10,7,7,10  \\
2M 1328$+$2114     & 13 28 55.04 & $+$21 14 48.6 & L5.0 & 20.09 & 3.90 &   1.92 & 0.06 & 0.11 & 0.08 &   31.0 &   3.8  & 14,10,7,7,10  \\
S* 1346$-$0031     & 13 46 46.45 & $-$00 31 50.4 & T6.0 & 21.03 & 5.27 &   0.11 & 0.18 & 0.05 & 0.05 &   72.7 &   5.0  & 19,10,15,15,8 \\
LHS 2924           & 14 28 43.23 & $+$33 10 39.1 & M9.0 & 15.21 & 3.22 &   1.25 & 0.02 & 0.02 & 0.02 &   90.8 &   1.3  & 11,23,7,7,24  \\
2M 1439$+$1929     & 14 39 28.37 & $+$19 29 15.0 & L1.0 & 16.12 & 3.36 &   1.21 & 0.01 & 0.02 & 0.02 &   69.6 &   0.5  & 14,10,7,7,10  \\
Gl 570D            & 14 57 15.00 & $-$21 21 48.0 & T8.0 & 20.94 & 5.90 &$-$0.39 & 0.18 & 0.05 & 0.05 &  169.3 &   1.7  & 19,10,15,15,20\\
T* 513-46546       & 15 01 08.19 & $+$22 50 02.0 & M9.0 & 15.16 & 3.29 &   1.16 & 0.04 & 0.02 & 0.02 &   94.4 &   0.6  & 11,10,7,7,10  \\
2M 1507$-$1627     & 15 07 47.68 & $-$16 27 41.0 & L5.0 & 16.69 & 3.82 &   1.56 & 0.11 & 0.08 & 0.07 &  136.4 &   0.6  & 9,2,2,2,10    \\
T* 868-110639      & 15 10 16.86 & $-$02 41 07.4 & M9.0 & 15.73 & 3.09 &   1.32 & 0.05 & 0.07 & 0.09 &   57.5 &   1.9  & 11,2,2,2,3    \\
Gl 584C            & 15 23 22.63 & $+$30 14 56.2 & L8.0 & 20.27 & 4.18 &   1.71 & 0.11 & 0.05 & 0.05 &   53.7 &   1.2  & 5,10,15,15,20 \\
S* 1624$+$0029     & 16 24 14.37 & $+$00 29 15.6 & T6.0 & 20.88 & 5.41 &$-$0.06 & 0.18 & 0.05 & 0.05 &   91.5 &   2.3  & 19,10,15,15,10\\
2M 1632$+$1904     & 16 32 29.11 & $+$19 04 40.7 & L8.0 & 19.98 & 4.11 &   1.87 & 0.05 & 0.05 & 0.05 &   65.6 &   2.1  & 14,10,7,7,10  \\
2M 1658$+$7027     & 16 58 03.81 & $+$70 27 01.6 & L1.0 & 16.68 & 3.39 &   1.37 & 0.02 & 0.02 & 0.02 &   53.9 &   0.7  & 12,10,7,7,10  \\
Epsilon Indi Ba    & 22 04 10.97 & $-$56 47 00.4 & T1.0 & 17.14 & 4.58 &   1.26 & 0.12 & 0.10 & 0.10 &  275.8 &   0.7  & 29,2,2,2,20   \\
Epsilon Indi Bb    & 22 04 10.97 & $-$56 47 00.4 & T6.0 & 18.64 & 5.14 &   0.02 & 0.12 & 0.10 & 0.10 &  275.8 &   0.7  & 29,2,2,2,20   \\
2M 2224$-$0158     & 22 24 43.85 & $-$01 58 52.1 & L4.5 & 18.03 & 4.08 &   1.98 & 0.17 & 0.08 & 0.09 &   88.1 &   1.1  & 5,2,2,2,10    \\
2M 2356$-$1553     & 23 56 54.77 & $-$15 53 11.1 & T6.0 & 21.21 & 5.44 &   0.05 & 0.10 & 0.06 & 0.18 &   69.0 &   3.4  & 19,25,2,2,8   \\
    \noalign{\smallskip}
    \hline 
   \end{tabular}
 $$
  \begin{list}{}{}
  \item[]Abbreviations.---2M: 2MASS; T*: TVLM; S*: SDSS; D*: DENIS

Cols. (1)--(3): object name, right ascension and declination for equinox J2000 from SIMBAD;
col. (4): spectral type; cols. (5)--(7), and (8)--(10): $I$-magnitude, colours and associated standard errors;
cols. (11) \& (12): trigonometric parallax and its standard error;
col. (13): references for columns 4--7, and 11.\\
References.---(1): \citet{golimowski}; (2): DENIS photometry; (3): \citet*{altena95}; (4): \citet{geballe};
(5): \citet{kirk00}; (6): \citet{gizis03}; (7): 2MASS photometry; (8): \citet{vrba}; (9): \citet{reid00};
(10): \cite{dahn}; (11): \citet{martin99}; (12): \citet{gizis00}; (13): \citet*{kirk97}; (14): \citet{kirk99};
(15): \citet{stephens}; (16): \citet{tinney96}; (17): \citet{leggett01}; (18): \citet{fan}; (19): \citet{bur02};
(20): HIPPARCOS; (21): \citet{close03}; (22): \citet{reid01}; (23): \citet*{leggett98}; (24): \citet{monet};
(25): \citet{bur03}; (26): \citet{del01}; (27): \citet{deacon01}; (28): \citet{koerner}; (29): \citet{mccau}. 

  \end{list}
\end{table*}

\begin{table*}
  \caption{29 DENIS ultracool dwarf candidates}
  \label{candidates}
  $$
  \begin{tabular}{llllllllll}
   \hline 
   \hline
   \noalign{\smallskip}
 DENIS-P name          &  $\alpha_{2000}$ &  $\delta_{2000}$ &  Epoch           &
 $I$                 &  $I-J$           &  $J-K$           &  err$I$          &
 err$J$              &  err$K$              \\
   \hline
   \noalign{\smallskip}  
J0615493$-$010041 & 06 15 49.32 & $-$01 00 41.2 & 1996.060 & 17.00 & 3.24 & 1.45 & 0.10 & 0.08 & 0.11 \\
J0630014$-$184014 & 06 30 01.43 & $-$18 40 14.7 & 1999.863 & 15.88 & 3.17 & 1.33 & 0.06 & 0.07 & 0.08 \\
J0644143$-$284141 & 06 44 14.34 & $-$28 41 41.9 & 1996.954 & 16.96 & 3.17 & 1.14 & 0.10 & 0.09 & 0.14 \\
J0649299$-$154104 & 06 49 29.98 & $-$15 41 04.0 & 1998.123 & 18.36 & 3.96 & 1.74 & 0.22 & 0.30 & 0.15 \\
J0652197$-$253450 & 06 52 19.73 & $-$25 34 50.5 & 2001.066 & 15.95 & 3.26 & 1.17 & 0.06 & 0.08 & 0.09 \\
J0716478$-$063037 & 07 16 47.88 & $-$06 30 37.5 & 1997.107 & 17.46 & 3.55 & 1.30 & 0.11 & 0.09 & 0.13 \\
J0719234$-$173858 & 07 19 23.44 & $-$17 38 58.8 & 1996.948 & 18.10 & 3.76 & 2.00 & 0.19 & 0.11 & 0.11 \\
J0719358$-$174910 & 07 19 35.86 & $-$17 49 10.4 & 1998.978 & 18.06 & 3.52 & 1.87 & 0.16 & 0.11 & 0.14 \\
J0751164$-$253043 & 07 51 16.44 & $-$25 30 43.3 & 1999.189 & 16.53 & 3.31 & 1.15 & 0.07 & 0.07 & 0.11 \\
J0805110$-$315811 & 08 05 11.03 & $-$31 58 11.8 & 2000.060 & 16.50 & 3.29 & 0.84 & 0.09 & 0.08 & 0.13 \\
J0812316$-$244442 & 08 12 31.69 & $-$24 44 42.1 & 1996.301 & 17.27 & 3.38 & 1.39 & 0.12 & 0.10 & 0.14 \\
J0823031$-$491201 & 08 23 03.17 & $-$49 12 01.3 & 1999.973 & 17.14 & 3.56 & 1.72 & 0.11 & 0.09 & 0.09 \\
J0828343$-$130919 & 08 28 34.34 & $-$13 09 19.9 & 1996.109 & 16.09 & 3.37 & 1.40 & 0.06 & 0.07 & 0.10 \\
J1048278$-$525418 & 10 48 27.82 & $-$52 54 18.4 & 2001.359 & 17.25 & 3.26 & 1.45 & 0.13 & 0.11 & 0.14 \\
J1126399$-$500355 & 11 26 39.93 & $-$50 03 55.3 & 1999.263 & 17.80 & 3.85 & 1.21 & 0.15 & 0.09 & 0.13 \\
J1157480$-$484442 & 11 57 48.08 & $-$48 44 42.6 & 1996.246 & 17.41 & 3.38 & 1.41 & 0.12 & 0.09 & 0.13 \\
J1159274$-$524718 & 11 59 27.42 & $-$52 47 18.7 & 1999.389 & 14.49 & 3.12 & 1.08 & 0.03 & 0.07 & 0.07 \\
J1232178$-$685600 & 12 32 17.80 & $-$68 56 00.7 & 1999.167 & 15.43 & 3.05 & 1.08 & 0.05 & 0.06 & 0.07 \\
J1253108$-$570924 & 12 53 10.87 & $-$57 09 24.9 & 2000.216 & 16.74 & 3.29 & 1.42 & 0.09 & 0.08 & 0.10 \\
J1347590$-$761005 & 13 47 59.07 & $-$76 10 05.5 & 1999.189 & 17.05 & 3.24 & 1.38 & 0.11 & 0.09 & 0.11 \\
J1454078$-$660447 & 14 54 07.84 & $-$66 04 47.1 & 1998.375 & 16.90 & 3.70 & 1.48 & 0.10 & 0.07 & 0.09 \\
J1519016$-$741613 & 15 19 01.62 & $-$74 16 13.9 & 1996.322 & 16.56 & 3.07 & 0.92 & 0.07 & 0.08 & 0.12 \\
J1520022$-$442242 & 15 20 02.24 & $-$44 22 42.2 & 1999.301 & 16.69 & 3.45 & 1.42 & 0.08 & 0.07 & 0.09 \\
J1705474$-$544151 & 17 05 47.41 & $-$54 41 51.2 & 2000.301 & 16.66 & 3.20 & 1.27 & 0.09 & 0.09 & 0.11 \\
J1733423$-$165449 & 17 33 42.31 & $-$16 54 49.8 & 1996.369 & 16.82 & 3.21 & 1.17 & 0.10 & 0.09 & 0.11 \\
J1745346$-$164053 & 17 45 34.67 & $-$16 40 53.7 & 2000.552 & 17.11 & 3.37 & 1.36 & 0.12 & 0.11 & 0.12 \\
J1756296$-$451822 & 17 56 29.64 & $-$45 18 22.4 & 1999.321 & 15.46 & 3.14 & 1.13 & 0.06 & 0.07 & 0.08 \\
J1756561$-$480509 & 17 56 56.19 & $-$48 05 09.5 & 2000.533 & 16.74 & 3.41 & 1.19 & 0.12 & 0.10 & 0.11 \\
J1909081$-$193748 & 19 09 08.17 & $-$19 37 48.2 & 2001.466 & 17.92 & 3.58 & 1.72 & 0.17 & 0.12 & 0.12 \\
    \noalign{\smallskip}
    \hline 
   \end{tabular}
  $$
\end{table*}

\section{SELECTION CRITERIA}
We systematically searched the DENIS database (available at 
the Paris Data Analysis Center, PDAC) for potential 
members of the solar neighbourhood, with simple and 
well defined criteria. We first selected all DENIS 
objects which matched $|b|~<~15\degr$ (low 
Galactic latitude) and $I-J \geq 3.0$ (spectral type 
later than nominally M8.0, \citealt{leggett92}).
We then required that the position of the candidates 
in the ($I-J$, $J-K$) colour-colour diagram 
be within $J-K = \pm0.5$~mag of our linear fit 
to the locus of ultracool dwarfs (Fig.~\ref{IJ_JK}).
We then computed photometric distances using 
the ($M_{\rm J}$, $I-J$) relation established in Section~2 and retained the 
candidates with $d_{\rm phot} \leq 30$~pc. 

Before setting out to measure the -- labour intensive -- 
proper motions needed to use reduced proper motions 
as a discriminant between nearby dwarfs and distant 
giants (see Section~5 for further details), we eliminated the bright candidates 
($I<13.0$, $d_{\rm phot}<<30$~pc) 
within the boundaries of known low Galactic latitude molecular clouds
in SIMBAD. 
This left 29 ultracool dwarf candidates 
(Table~\ref{candidates}) for which we needed to 
measure proper motions. Their colour inferred spectral types range
from M8.0 to $\sim$L8.0 (Fig.~\ref{IJ_JK}). 
Six of these 29 candidates are previously known ultracool dwarfs:
2MASS J0644143$-$284141 (M8, \citealt{cruz03}); 
SSSPM J0829$-$1309 (L2, \citealt{scholz02}); 2MASS J1126399$-$500355 (L9, \citealt{folkes});
1RXS J115928.5$-$524717 (M9, \citealt{fuhr03, ham04}); 2MASS J1347590$-$761005 (L0, \citealt{kendall});
2MASSS J1520022$-$442242 (L1+L3.5, \citealt{kendall}; or L1.5+L4.5, \citealt{burl}).
We also searched for T dwarf candidates by using the ($I-J$, $J-K$) colour-colour diagram
along with a linear fit to the colours of T dwarfs from Table~\ref{cali},
however no T dwarf candidates were found. This reflects the fact that
the DENIS survey with limiting magnitudes of
$I=18.5$, $J=16.5$, $K=14.0$ is not very sensitive for T dwarf detections.
For example, a T0.5 dwarf ($M_{\rm I} \sim 19.3$, $M_{\rm J} \sim 14.8$ and
$M_{\rm J} \sim 13.2$, \citealt{vrba}) could be detected by DENIS in all three $I$, $J$ and $K$ bands
were its distance within $\sim$7~pc, e.g. Epsilon~Indi~B
\citep{scholz03}. A relaxation in the request for detections in all three DENIS bands, for example requiring
detections in the $J$ and $K$ bands but not the $I$ band may allow us to detect 
nearby T dwarfs; however we did not consider such a search in this paper.
\section{PROPER MOTION MEASUREMENTS}
We queried ALADIN 
\footnote{http://aladin.u-strasbg.fr/java/nph-aladin.pl}
and the Digital Sky Survey (DSS)
server\footnote{http://archive.stsci.edu/cgi-bin/dss\_plate\_finder} 
for publically available scanned plates containing 
these candidates. Both ALADIN and the DSS archive provide
access to the plates of the POSS-I ($R$-band), SERC-R 
($R$-band), SERC-I ($I$-band), POSS-II F ($R$-band), and POSS-II N 
($I$-band) surveys. ALADIN additionally contains digitized images
of the ESO-R survey ($R$-band), which often extend the 
time baseline enough to significantly reduce the standard error 
of our proper motion measurements. We used SExtractor 
\citep{bertin} to extract coordinates of our targets from 
all available digitized images. We then determined proper 
motions through a least-square fit to the positions at the 3 
to 5 available epochs with time baselines spaning from 10 to 21 years, 
including DENIS\footnote{http://cdsweb.u-strasbg.fr/denis.html} 
and 2MASS\footnote{http://cdsweb.u-strasbg.fr/viz-bin/VizieR}.
The uncertainty on the proper motion measurement is the rms dispersion
around that fit. 

Table~\ref{MLdwarfs} and \ref{redobj} present our 
proper motions and the associated standard errors
for 26 ultracool dwarfs and 3 reddened main sequence stars, respectively
(see Section~5 and 6 for further details).

\begin{table*}
 \caption{Twenty six nearby L, and late-M dwarfs}
\label{MLdwarfs}
 $$
 \begin{tabular}{lccrlrllllll}
   \hline 
   \hline
   \noalign{\smallskip}
 DENIS name  &   Time &  No. of   &$\mu$RA  &  err$\mu$RA &  $\mu$DE &  err$\mu$DE & 
 $H_{\rm I}$ &  err$H_{\rm I}$ &  $H_{\rm I}^{\rm max}$  & $d$         & err$d$   \\
             &  b.l. (yr) &  obs.  & (\arcsec yr$^{-1}$) &  (\arcsec yr$^{-1}$) &  (\arcsec yr$^{-1}$) & (\arcsec yr$^{-1}$) &
 (mag) &  (mag)  &  (mag) & (pc)  & (pc)  \\
 (1) & (2) & (3) & (4) & (5) & (6) & (7) & (8) & (9) & (10) & (11) & (12)  \\
   \hline
   \noalign{\smallskip}  
D* J0615493$-$010041 & 15.616 & 4 &  0.226 & 0.012 & $-$0.075 & 0.014 & 18.9 & 0.2 &  7.9 & 26.8 & 4.7  \\
D* J0630014$-$184014 & 11.792 & 4 &  0.350 & 0.008 & $-$0.503 & 0.004 & 19.8 & 0.1 &  7.9 & 18.0 & 3.1  \\
D* J0644143$-$284141 & 10.912 & 4 &  0.329 & 0.019 & $-$0.105 & 0.039 & 19.7 & 0.3 &  7.9 & 29.6 & 5.3  \\
D* J0652197$-$253450 & 21.151 & 3 & $-$0.233 & 0.005 &  0.086 & 0.002 & 17.9 & 0.1 &  7.9 & 16.0 & 2.8  \\
D* J0716478$-$063037 & 18.120 & 3 & $-$0.016 & 0.013 &  0.121 & 0.005 & 17.9 & 0.2 &  8.1 & 18.7 & 3.4  \\
D* J0751164$-$253043 & 15.123 & 3 & $-$0.885 & 0.003 &  0.142 & 0.004 & 21.3 & 0.1 &  7.9 & 19.1 & 3.3  \\
D* J0805110$-$315811 & 21.024 & 3 & $-$0.231 & 0.004 &  0.103 & 0.010 & 18.5 & 0.2 &  7.9 & 19.6 & 3.4  \\
D* J0812316$-$244442 & 18.899 & 3 &  0.096 & 0.001 & $-$0.165 & 0.007 & 18.7 & 0.2 &  8.0 & 23.7 & 4.4  \\
D* J0823031$-$491201 & 20.994 & 3 & $-$0.137 & 0.005 &  0.017 & 0.001 & 17.8 & 0.2 &  8.1 & 15.9 & 2.9  \\
D* J0828343$-$130919 & 14.293 & 4 & $-$0.547 & 0.018 &  0.078 & 0.009 & 19.8 & 0.1 &  8.0 & 14.0 & 2.4  \\
D* J1048278$-$525418 & 21.053 & 3 & $-$0.179 & 0.009 &  0.033 & 0.017 & 18.6 & 0.3 &  7.9 & 29.1 & 5.5  \\
D* J1126399$-$500355 & 14.153 & 3 & $-$1.570 & 0.000 &  0.438 & 0.011 & 23.9 & 0.2 &  8.4 & 12.5 & 2.2  \\
D* J1157480$-$484442 & 16.178 & 3 & $-$0.052 & 0.016 &  0.001 & 0.001 & 16.0 & 0.8 &  8.0 & 25.3 & 4.5  \\
D* J1159274$-$524718 & 19.154 & 5 & $-$1.067 & 0.003 & $-$0.119 & 0.026 & 19.6 & 0.0 &  7.8 & 10.2 & 1.7  \\
D* J1232178$-$685600 & 15.890 & 3 & $-$0.196 & 0.010 & $-$0.068 & 0.008 & 17.0 & 0.2 &  7.8 & 17.4 & 2.9  \\
D* J1253108$-$570924 & 20.033 & 4 & $-$1.575 & 0.001 & $-$0.435 & 0.016 & 22.8 & 0.1 &  7.9 & 21.8 & 3.8  \\
D* J1347590$-$761005 & 19.836 & 4 &  0.193 & 0.009 &  0.049 & 0.020 & 18.5 & 0.3 &  7.9 & 27.5 & 4.9  \\
D* J1454078$-$660447 & 19.054 & 3 &  0.525 & 0.002 & $-$0.376 & 0.006 & 21.0 & 0.1 &  8.3 & 10.9 & 1.9  \\
D* J1519016$-$741613 & 15.781 & 3 &  0.317 & 0.010 &  0.097 & 0.009 & 19.2 & 0.1 &  7.8 & 28.5 & 5.0  \\
D* J1520022$-$442242 & 20.978 & 4 & $-$0.623 & 0.008 & $-$0.390 & 0.016 & 21.0 & 0.1 &  8.0 & 15.9 & 2.7  \\
D* J1705474$-$544151 & 20.879 & 4 & $-$0.072 & 0.008 &  0.030 & 0.001 & 16.1 & 0.3 &  7.9 & 24.5 & 4.4  \\
D* J1733423$-$165449 & 17.905 & 3 &  0.081 & 0.015 & $-$0.048 & 0.015 & 16.7 & 0.6 &  7.9 & 26.0 & 4.7  \\
D* J1745346$-$164053 & 17.914 & 4 &  0.116 & 0.005 & $-$0.111 & 0.019 & 18.1 & 0.3 &  8.0 & 22.4 & 4.2  \\
D* J1756296$-$451822 & 19.916 & 5 &  0.064 & 0.005 & $-$0.183 & 0.006 & 16.9 & 0.1 &  7.8 & 15.5 & 2.6  \\
D* J1756561$-$480509 & 13.045 & 4 &  0.078 & 0.004 &  0.050 & 0.011 & 16.6 & 0.3 &  8.0 & 17.6 & 3.2  \\
D* J1909081$-$193748 & 14.776 & 5 & $-$0.064 & 0.021 & $-$0.145 & 0.034 & 18.9 & 0.7 &  8.1 & 21.9 & 4.2  \\
    \noalign{\smallskip}
    \hline
   \end{tabular}
 $$
\begin{list}{}{}
\item[]Abbreviations.---D*: DENIS-P \\
Col. (1): abbreviated DENIS name.
Cols. (2) \& (3): time baseline and number of observations.
Cols. (4)--(7): proper motions and associated errors.
Cols. (8) \& (9): $I$-band reduced proper motion and associated error.
Col. (10): maximum reduced proper motion for an M giant of the same $I-J$ colour.
Cols. (11) \& (12): photometric distance and its associated error.
\end{list}
\end{table*}
\begin{table*}
  \caption{Three red, distant objects}
\label{redobj}
  $$
 \begin{tabular}{lclrlrllll}
   \hline 
   \hline
   \noalign{\smallskip}
 DENIS name  &  Time &  No. of   &$\mu$RA  &  err$\mu$RA &  $\mu$DE &  err$\mu$DE & 
 $H_{\rm I}$ &  err$H_{\rm I}$ &  $H_{\rm I}^{\rm max}$ \\
             &  b.l. (yr) &  obs.  & (\arcsec yr$^{-1}$) &  (\arcsec yr$^{-1}$) &  (\arcsec yr$^{-1}$) & (\arcsec yr$^{-1}$) &
 (mag) &  (mag)  &  (mag) \\
 (1) & (2) & (3) & (4) & (5) & (6) & (7) & (8) & (9) & (10) \\
   \hline
   \noalign{\smallskip}  
D* J0649299$-$154104 & 15.860 & 4 &  0.024 & 0.021 & $-$0.033 & 0.072 & 16.4 & 4.0 &  8.6  \\
D* J0719234$-$173858 & 16.179 & 4 &  0.006 & 0.049 & $-$0.011 & 0.009 & 13.6 & 5.6 &  8.3  \\
D* J0719358$-$174910 & 16.899 & 4 & $-$0.007 & 0.013 & $-$0.006 & 0.008 & 12.9 & 3.7 &  8.1  \\
    \noalign{\smallskip}
    \hline 
   \end{tabular}
   $$
  \begin{list}{}{}
  \item[] 
   Cols. (1)--(10): same as Table~\ref{MLdwarfs}.\\
  \end{list}
\end{table*}
\begin{table}
  \caption{Intrinsic colors for giants}
\label{m_giants}
  $$
 \begin{tabular}{lllrl}
   \hline 
   \hline
   \noalign{\smallskip}
MK & $M_{\rm I}$ & $I-J$ & $H_{\rm I}^{\rm max}$ & References \\
   \hline
   \noalign{\smallskip}  
M4 & $-$3.49 & 1.38 & 7.7  & B88; T90 \\
M5 & $-$3.56 & 1.86 & 7.6  & B88; T90  \\
M6 & $-$3.59 & 2.28 & 7.6  & B88; T90 \\
M7 & $-$3.22 & 2.75 & 7.9  & B88; T90  \\
M8 & $-$2.96 & 3.67 & 8.2  & \cite{fluks} \\
M8 & $-$0.88 & 4.79 & 10.3 & \cite{fluks}  \\
    \noalign{\smallskip}
    \hline 
   \end{tabular}
   $$
\end{table}
\begin{figure}
\psfig{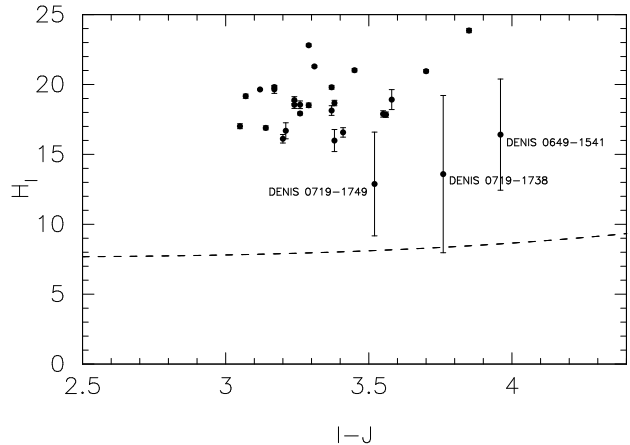}
\caption{$I$-band reduced proper motions versus $I-J$. The dashed 
curve represents the maximum possible reduced proper motion for
a giant, $H_{\rm I}^{\rm max}$. Objects above this curve must be
dwarfs. 
Three red, distant dwarfs are indicated:
DENIS~0649$-$1541, DENIS~0719$-$1738 and DENIS~0719$-$1749
\label{HI_IJ}}
\end{figure}
\begin{figure}
\psfig{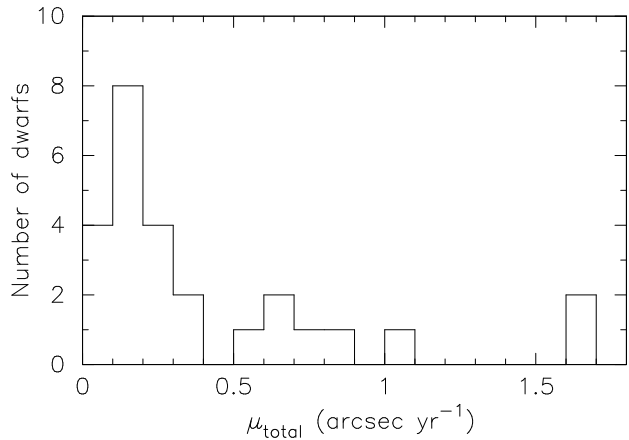}
\caption{Proper motion distribution for the 26 ultracool dwarfs M8--L5.5 in this paper.
The bin size is 0.1$\arcsec$yr$^{-1}$.}
\label{distribution_pm}
\end{figure}
\section{L AND LATE-M DWARF CANDIDATES CLASSIFIED BY MAXIMUM 
REDUCED PROPER MOTION}

As in our search for nearby mid-M dwarfs \citep{p03}, 
we use the MRPM method to reject giants. This method uses the 
reduced proper motion 
($H=M+5\log(V_{\rm t}/4.74)$ = $m+5+\log({\mu})$) 
versus colour diagram, where nearby ultracool dwarfs and distant 
giants segregate very cleanly. Concretely, we use the 
colour-magnitude relation of giants to compute the 
maximum reduced proper motion that a red giant can have at 
a given colour -- setting its tangential velocity to the
Galactic escape velocity, for which we adopt a conservative 
800~km~s$^{-1}$ (see \citealt{p03} and references therein) --, 
and then declare any object with a reduced proper
motion above that value for its colour to be a dwarf. The details
of the MRPM method are given in \citet{p03}.

To use this method here for the cooler M8.0--L8.0 candidates, we 
had to extend the maximum reduced proper motion 
curve of giants towards redder colours ($I-J \geq 3.0$), 
adding cooler giants from \citet{fluks} to the
\citet{the} (hereafter T90); \citet{bessell88} (hereafter B88) 
sample used in our earlier article.  Table~\ref{m_giants} presents
the photometric data used in this paper and the maximum reduced proper motion
computed for giants $H_{\rm I}^{\rm max}$ at the given colors.
The following cubic least-square fit (see 
Fig.~\ref{HI_IJ}) is valid for 
$3.0 \leq I-J \leq 4.5$, or dwarf spectral 
types between M8.0 and L8.0:

\begin{eqnarray}
   H_{\rm I}^{\rm max} & = & 6.79 + 1.25(I-J)-0.630(I-J)^{2} \nonumber \\
                 &   & +0.108(I-J)^{3} \label{eq2}                 
\end{eqnarray}
The rms dispersion around the fit is 0.1~mag.
Figure~\ref{HI_IJ} shows the position of the resulting 
$H_{\rm I}^{\rm max}$~vs.~($I-J$) curve relative to
the 29 candidates. The curve well classifies 26 candidates as
late-M and L dwarfs (Table~\ref{MLdwarfs}). The likely giants are
DENIS~0649$-$1541 and DENIS~0719$-$1749 (further discussed in Section~6) 
that are in the ultracool dwarf part of the diagram but within 2$\sigma$ of the 
limit, and the remaining one DENIS~0719$-$1738 that is within 1$\sigma$. 

Figure~\ref{distribution_pm} presents the proper motion distribution
of our sample of 26 ultracool dwarfs (Table~\ref{MLdwarfs}). 
The sample consists of
22 high-pm ($\mu \geq 0.1\arcsec$yr$^{-1}$) objects and 4 low-pm ($\mu < 0.1\arcsec$yr$^{-1}$) ones, 
giving a low-pm object fraction of 15.4$\%$
consistent with one estimated for mid-M dwarfs (M5--M8) of 14.8$\%$ 
(see \citealt{crifo,pb}). The finding charts of the 26 ultracool dwarfs
are given in Fig.~\ref{finding_charts}, 20 of these 26 dwarfs are new,
6 are already known in the literature (see the caption of Table~6 for references).

\begin{figure*}
\caption{$I$-band DENIS finding charts for the 26 L and late-M dwarfs listed in Table~\ref{MLdwarfs}. 
The charts are 4.0\arcmin$\times$4.0\arcmin, with North
is up and East is to the left. [The finding charts will appear here, this figure ($>$3 Mb) is removed
in the version for astro-ph]
\label{finding_charts}}
\end{figure*}
%
%
%\setcounter{figure}{4}
%\begin{figure*}
%\psfig{width=19.0cm,file=f5b.ps,angle=0}
%\caption{continued.
%\label{finding_charts_b}}
%\end{figure*}
%
%\setcounter{figure}{4}
%\begin{figure*}
%\psfig{width=19.0cm,file=f5c.ps,angle=0}
%\caption{continued.
%\label{finding_charts_c}}
%\end{figure*}
%
\begin{figure*}
\psfig{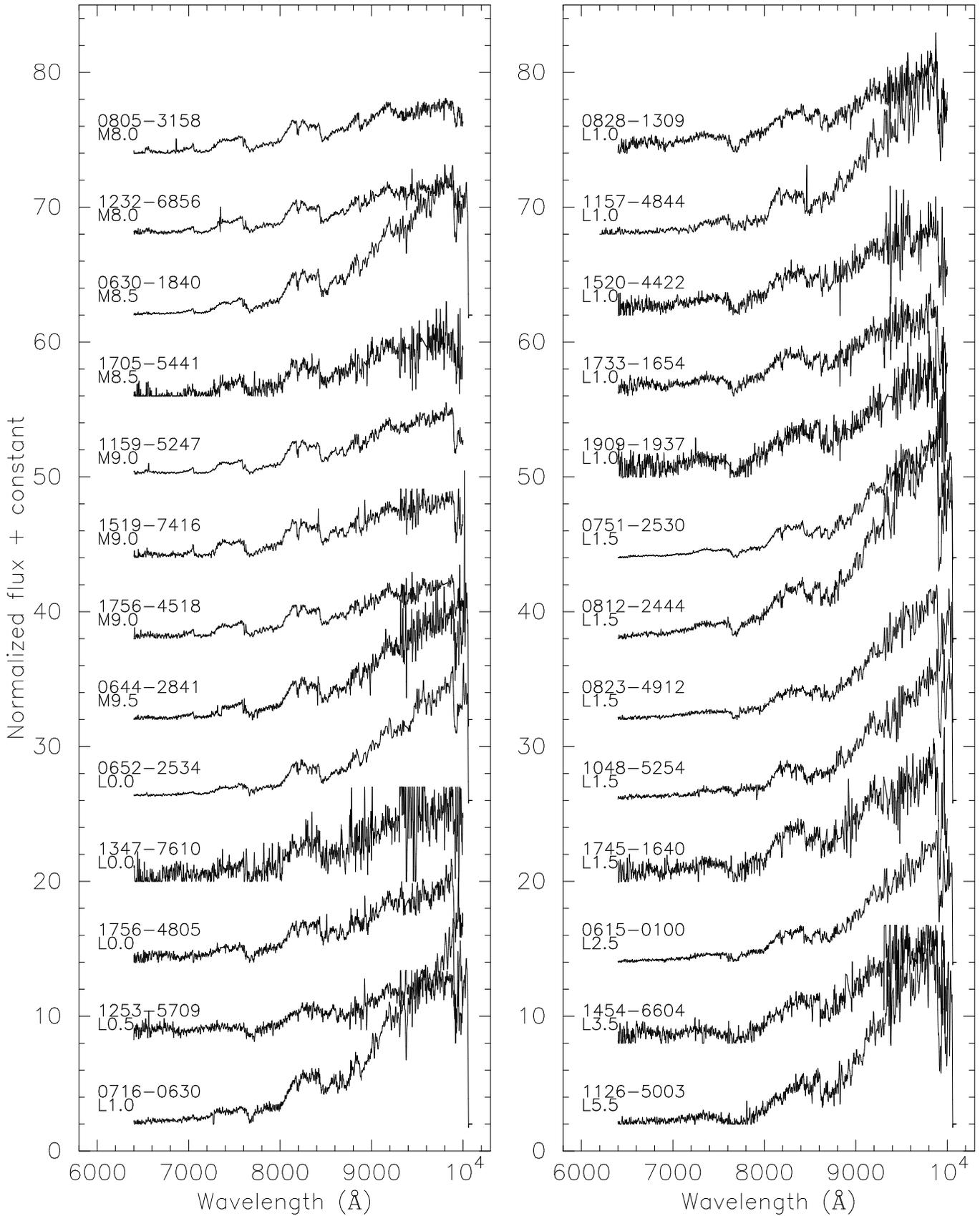}
\caption{Spectra of the 26 M8--L5.5 ultracool dwarfs are plotted by increasing spectral types. 
\label{spectra_dwarfs}}
\end{figure*} 
\section{Spectroscopic observations and analysis}
\subsection{Optical spectroscopic observations}
We observed the candidates on December 17--18 2006 and on May 20-25 2007
with the Double Beam Spectrograph on the ANU (Australian National University) 
2.3m telescope at Siding Spring Observatory.  
Both gratings 158g/mm and 316g/mm were used and they 
provided a wavelength coverage of 6100--10200~\AA~at 5 and 2~\AA~resolution 
for gratings 158g/mm and 316g/mm, respectively.
Exposures of 600--1800 seconds were taken, depending on the target
magnitude. The seeing was around 2.0\arcsec.
The data were reduced using 
the FIGARO\footnote{http://www.aao.gov.au/AAO/figaro/} data reduction system. 
Smooth spectrum stars were 
observed at a range of airmass to remove the telluric lines using the technique
of \citet{bessell99}. The EG131 \citep{bessell99} spectrophotometric standard 
was used for relative flux calibration, and a NeAr arc provided the 
wavelength calibration. All spectra were normalized over the 7540--7580~\AA~interval, 
the denominator of the PC3 index \citep{martin99} and a region 
with a good flat pseudo-continuum. 
We also observed VB~8 (M7, for spectral classification, see \citealt{martin99}), 
VB~10 (M8, \citealt{martin99}), 
LHS 2065 (M9, \citealt{martin99}), 2MASS~J22431696$-$5932206 (L0, \citealt{kendall}),
2MASS~J01282664$-$5545343 (L1, \citealt{kendall}) and 
2MASS~J09211410$-$2104446 (L2, \citealt{reid06}) with the same instrument setup
and used them as templates.

At the resolution of the spectra, M and L dwarfs are immediately distinguished
from M giants by the presence of the Na{\small I} and K{\small I} doublets, 
the presence of FeH bands, the appearance of strong CaH cutting into the 
continuum shortward of 7000~\AA, and by the absence of the Ca{\small II} triplet 
(e.g. \citealt{bessell91}).

\begin{figure}
\psfig{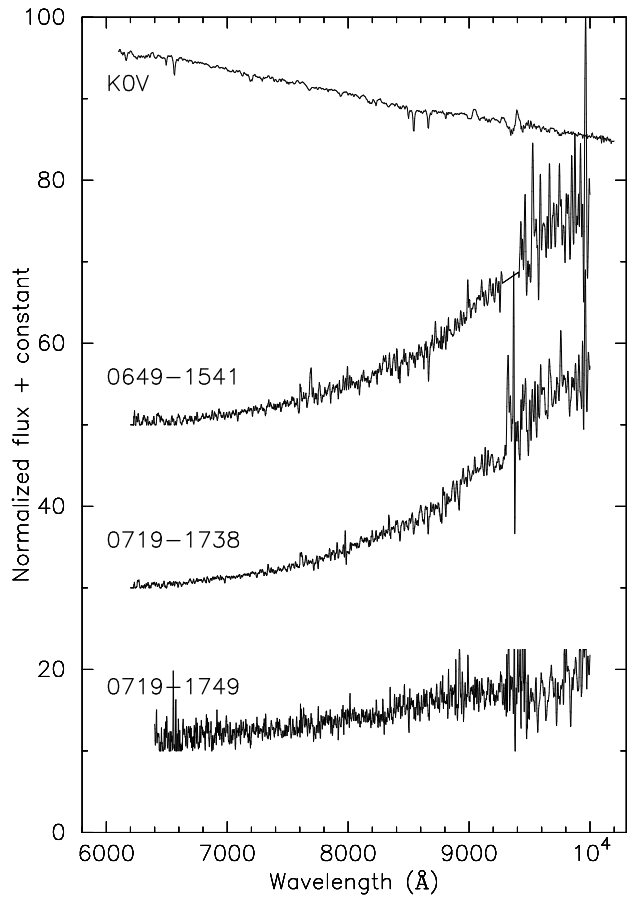}
\caption{Spectra of three reddened F-G main sequence stars (Table~\ref{redobj}) ({\it bottom})
A K0V dwarf from \citet{pickles} is also plotted for comparison ({\it top}).
\label{spectra_redobj}}
\end{figure}

Figure~\ref{spectra_dwarfs} presents spectra of 26 ultracool dwarfs (Table~\ref{MLdwarfs}), and
Figure~\ref{spectra_redobj} shows spectra of the 3 reddened F-K main sequence stars:
DENIS~0649$-$1541, DENIS~0719$-$1738, and DENIS~0719$-$1749. 
These three stars are probably in the background of some molecular
clouds and reddened by intervening dust. 
Using SIMBAD, we found that DENIS~0649$-$1541 was in the background of the 
[KKY2004]~G226.7$-$07.5 molecular cloud, while DENIS~0719$-$1738 and 
DENIS~0719$-$1749 
were in the vicinity of the ESO~559$-$6 planetary nebula.

\subsection{Spectral type classifications and distances}
To estimate the spectral types of L and late-M dwarfs, we used the PC3 index defined in 
\citet{martin99} and TiO5 defined in \citet*{reid95}. The VOa index, defined 
in \citet{kirk99}, saturates in the M8-L0.5 spectral type range
which covers a significant number of our ultracool dwarfs, we therefore 
only used it as a reference when there was a large difference (e.g. greater than 
2 subclasses) in the spectral type estimate between PC3 and TiO5.
We used spectral type versus spectral index relations to estimate spectral types, 
the \citet{cruz02} relation for TiO5 and  the \citet{martin99} relation for PC3.
The uncertainty in the spectral indices was calculated based on the rms dispersion 
of the flux around the mean in the wavelength regions that were used to 
calculate the spectral indices. 

Table~6 (column 13) lists the average value of the two spectral types computed from
PC3 and TiO5,  
except for three dwarfs DENIS~1126$-$5003, DENIS~1705$-$5441, and DENIS~1909$-$1937 
whose spectral type estimates were based on the PC3 index only.

\begin{figure}
\psfig{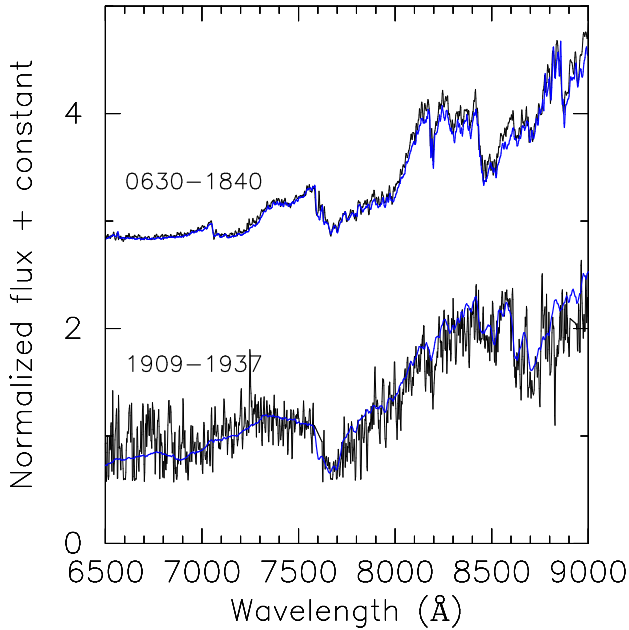}
\caption{Comparison of spectra between our objects (black lines) and templates (blue lines):
DENIS 0630$-$1840 and our M8 (VB 10, \citealt{vanB}) template ({\it top}), 
DENIS 1909$-$1937 and an L2 template from \citep{martin99} ({\it bottom}).
\label{spectra_comp}}
\end{figure}

We also compared all spectra with our templates 
and with others from \citet{martin99}. The \citet{martin99} template spectral type
ranged from M8 down to L6. Spectral type estimates from the best fits are listed
in Table~6 (column~15) with a typical uncertainty of 1.0 subclass.
These values are in good agreement with our spectral types estimated from spectral indices.
Figure~\ref{spectra_comp} shows two representative cases: DENIS~0630$-$1840 and
DENIS~1909$-$1937. The final spectral type is an average value
of the spectral type estimated from spectral indices (column 13, Table 6) and
one from the best fit of templates (column 15, Table 6) with an uncertainty of 1.0 subclass.

To estimate the absolute magnitude $M_{\rm J}$ of the ultracool dwarfs,
we used the absolute magnitude $M_{\rm J}$ versus spectral type relation
derived from 34 late-M, and L dwarfs (Table~\ref{cali}). 
The T dwarfs were excluded here since no T dwarf candidates had been found 
in our search.
A linear least-squares fit (Fig.~\ref{Mj_SpT})
to the data gives the following absolute magnitude-spectral type relation: 
$M_{\rm J} = 8.19 + 0.364 \times$SpT, where SpT is the spectral subtype,
counted from 8.0 for spectral type M8.0 to 18.0 for spectral type L8.0. 
The standard rms is 0.30~mag, corresponding to a 14\% error on distances.
One should note that for the spectral type range M8--L5.5 of our 26 ultracool dwarfs, the rms
is 0.27~mag which is somewhat higher than that of \citet{dahn} (0.25~mag) 
but smaller than the 0.37~mag of \citet{liu} and 0.40~mag of \citet{knapp04}.

Table~\ref{dist} lists our spectrophotometric distances
and associated errors. The errors were computed from an uncertainty of $\sim$0.37~mag on $M_{\rm J}$
corresponding to a spectral type uncertainty of 1.0 subclass and errors in
the apparent $J$-band magnitudes.

One should also note that some of the 26 ultracool dwarfs must be unresolved binaries, 
whose distances are underestimated by up to $\sqrt{2}$. 
\citet{kendall} recently reported that DENIS~1520$-$4422
(as 2MASSJ~15200224$-$4422419, see also \citealt{burl}) is an L1+L3.5 binary. According to
\citet{bouy03}, we can expect the binary fraction of our sample to be about 10$\%$.

Table~\ref{dist} lists our tangential velocity computed
from the proper motion and the spectrophotometric distance. 
Two of the dwarfs have relative high $V_{\rm t}$:
150 and 82~km~s$^{-1}$ for DENIS~1253$-$5709 and DENIS~1126$-$5003, respectively,
and suggests they may be mild subdwarfs like Kapteyn's star. A comparison of their 
spectra with the ultracool subdwarf spectra \citep*{bessell82,bur07} clearly excludes 
the possibility that they are ultracool extreme subdwarfs.
In addition, all ultracool subdwarfs ($I-J~\geq~2.0$, later than $\sim$sdM9) known to
date have $J-K<0.7$ (e.g. Table~7 
of \citealt{bur07}, and references therein) meanwhile both 
DENIS~1253$-$5709 and DENIS~1126$-$5003 have $J-K>1.2$. 
\citet{folkes} have recently reported that DENIS~1126$-$5003 (or 2MASS~J11263991$-$5003550)
is a possible L-T transition object. This can explain the fact that the L dwarf has an 
optical spectral type of L6.0 (this paper) significantly earlier than its 
near-infrared spectral type of L9.0 \citep{folkes}. 
Further observations would be useful to 
confirm the nature of this object as discussed in detail in their paper.
\citet{tsuji05} discusses complications due to dust formation 
in the spectra of the L-T transition objects. 
\subsection{Chromospheric activity}
Using the IRAF\footnote{IRAF is distributed by the National Optical Astronomy Observatories,
    which are operated by the Association of Universities for Research
    in Astronomy, Inc., under cooperative agreement with the National
    Science Foundation.} 
task SPLOT, we measured H$\alpha$ equivalent widths of the 26 ultracool dwarfs
(Table~\ref{dist}, column 9). Only the M9 dwarf DENIS~1159$-$5247 (or 1RXS~J115928.5$-$524717) shows 
strong H$\alpha$ emission. This M9 dwarf has also shown strong X-ray flaring emission \citep{fuhr03,ham04} demonstrating that the dwarf has chromospheric and coronal activity. 
We measured an upper limit for the remaining dwarfs, which have
weak or no H$\alpha$ emission or low signal-to-noise spectra. Unsurprisingly,
the dissipation of chromospheric activity in almost our ultracool dwarfs is due to
their predominantly cool, dense, and highly neutral atmospheres \citep{meyer, mohanty02}.
\subsection{A wide binary candidate $\sim$M0+L0}
During the SERC-I ($I$-band, epoch = 1980.344) image examination, 
we discovered a bright star at the position 
$\alpha_{2000}$=13$^{\rm h}$47$^{\rm m}$55.82$^{\rm s}$, $\delta_{2000}$=$-$76$\degr$10$\arcmin$20.2$\arcsec$, 
which has a common proper motion with 
DENIS~1347$-$7610 (L0, $\mu$RA$=0.193\pm0.009\arcsec$yr$^{-1}$, $\mu$DE$=0.049\pm0.020\arcsec$yr$^{-1}$).
We measured its proper motion $\mu$RA$=0.181\pm0.011\arcsec$yr$^{-1}$,
and $\mu$DE$=0.031\pm0.025\arcsec$yr$^{-1}$, 
in good agreement with the proper motion of the L0
dwarf. This primary (hereafter ASIAAa) has DENIS photometry $I=9.67\pm0.02$, $J=8.65\pm0.05$, 
and $K=7.81\pm0.05$, giving an $I$-band reduced proper motion ($H_{\rm I} = I + 5\log\mu + 5$) of 11.0$\pm$0.2.
This value is well above ($\sim12\sigma$) the maximum reduced proper motion \citep{p03} that a red giant can have at the 
colour of ASIAAa ($I-J=1.02$) indicating it must be a dwarf.
Follow-up spectroscopic observation of ASIAAa confirmed that the star is indeed an early M dwarf.  

Figure~\ref{spectra_comp_M0} shows the optical spectrum at 2~\AA~resolution 
of ASIAAa observed at the CTIO 4m-telescope on July 17 2007. It is very similar to 
the spectrum of HD~111631 ($\sim$M0) from 
MILES\footnote{http://www.ucm.es/info/Astrof/miles/database/database.html}
(A Medium resolution INT Library of Empirical Spectra). We therefore adopt
a spectral type of M0 for ASIAAa, giving an $I$-band absolute magnitude $M_{\rm I} = 7.06$ \citep{bessell91}.
Comparison with the DENIS $I$-band apparent magnitude then gives a distance of 33.3$\pm$6.1~pc,
in good agreement with the spectrophotometric distance $d_{\rm sp} = 24.9\pm5.2$~pc
of the L0 dwarf DENIS~1347$-$7610 (hereafter ASIAAb).
The presence of Balmer line emission (H$\alpha$ and H$\beta$) in ASIAAa indicates that the star is also chromospherically active.
The projected angular separation between ASIAAa and ASIAAb is 16.8\arcsec or 418~AU at the distance of 24.9~pc.
\section{Sample completeness, and the local ultracool dwarf density} 

The spectral types of almost all our ultracool dwarfs range from M8 to L3.5, excepting
 DENIS~1126$-$5003 (L5.5). 
For an L3.5 dwarf, we have $M_{\rm J}=12.95$, $M_{\rm K}=11.46$ (e.g. \citealt{vrba}),
and $I-J \sim 3.7$ (see Table~\ref{cali}). 
The limiting magnitudes of DENIS are $I=18.5$, $J=16.5$, and $K=14.0$, giving
a detection limited distance for L3.5 dwarfs of $\sim$23.5~pc.
This limit for L5.5 dwarfs is about 15~pc, we therefore did not include 
DENIS~1126$-$5003 (L5.5) in our sample for the statistics. 
One should note that a relaxation of the request of detections in all three DENIS bands 
(e.g. only requiring detections in the $J$ and $K$ bands)
might allow us to detect cooler L dwarfs 
as we discussed for T dwarf detection (see Section 3). However such a search is
outside the scope of this paper.

The differential spectrophotometric distance distribution of that sample 
(Fig.~\ref{distribution_function}) is well fitted by a $d^2$ distribution,
as expected for a constant density population, out to $\sim$22~pc. The
difference from the initial 30~pc selection cutoff reflects
the detection limiting distance of DENIS for ultracool dwarfs and
the different colour-magnitude relations used in the selection and in the final 
spectrophotometric distance estimate. We conservatively adopt 22~pc as the 
completeness limit of our sample, and use the 18 objects (Table~\ref{dist}) within that
distance to determine the local density of ultracool dwarfs. 

A sample limited by spectrophotometric distance is effectively a magnitude-limited
sample with a spectral type (or colour)-dependent magnitude limit. 
As such, the dispersion in the spectral type-absolute magnitude relation
has two distinct effects on the statistics of the stellar population \citep*{stobie}.
First, the average luminosity of stars at a given spectral type (or colour) is increased; 
this is the classical Malmquist bias \citep{malm36}. Second, due to the dispersion of the distance
estimator a magnitude-limited sample includes more stars at a given spectral type.
\begin{figure}
\psfig{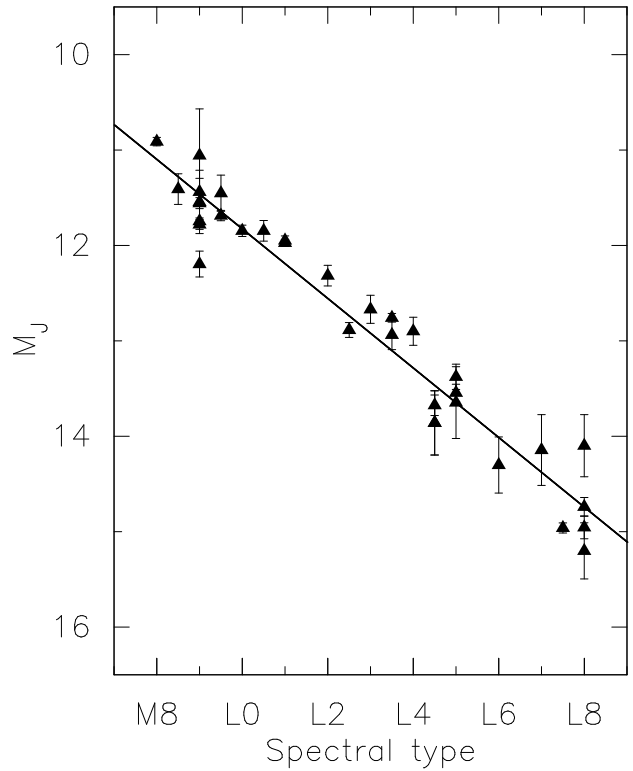}
\caption{Absolute magnitude $M_{\rm J}$ versus spectral type diagram for 34 late-M,
and L dwarfs, plotted as solid triangles (data in Table~\ref{cali}).
\label{Mj_SpT}}
\end{figure}
\begin{figure}
\psfig{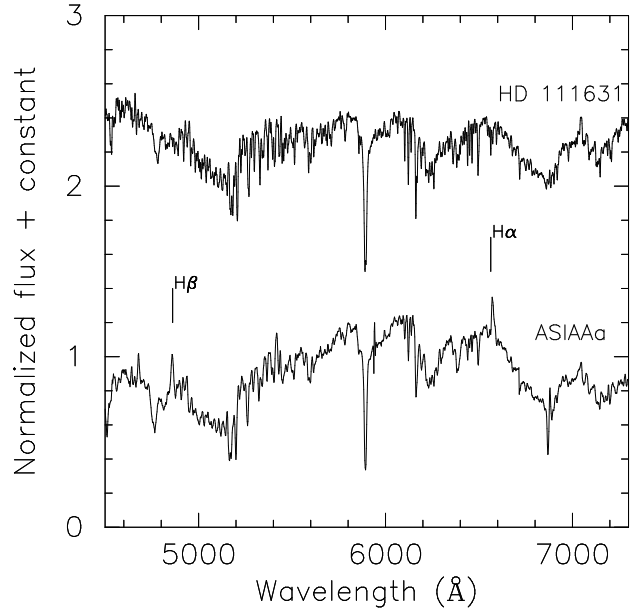}
\caption{Comparison of spectra between ASIAAa ({\it bottom}) and the HD~111631 (M0) template
({\it top}) from MILES. The Balmer line emission is indicated.
%\footnote{http://www.ucm.es/info/Astrof/miles/database/database.html}).
\label{spectra_comp_M0}}
\end{figure}

Here we study the sample of ultracool dwarfs M8--L3.5 over $11.1 \leq M_{\rm J} \leq 13.1$.
The first component 
of the Malmquist bias is therefore irrelevant, since firstly we do not look
for any significant luminosity resolution, and secondly the luminosity function 
is sufficiently flat over the M8--L3.5 spectral range \citep{cruz07}
that a small shift in the average luminosity will not
measurably affect the resulting density. The second component of the bias, 
on the other hand, is significant. For a gaussian dispersion of the 
colour-luminosity relation it can be computed analytically 
\citep{stobie}:
\begin{eqnarray}
\frac{{\Delta}{\Phi}}{\Phi}=\frac{1}{2}{\sigma}^2(0.6~{\ln}10)^2 \label{eq9}
\end{eqnarray}
\begin{figure}
\psfig{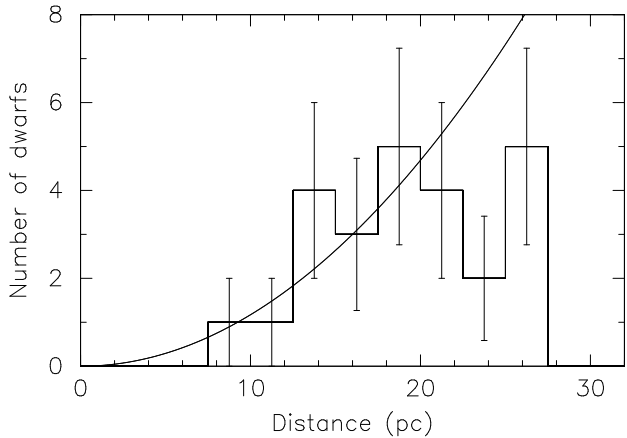}
\caption{Number of ultracool dwarfs per 2.5~pc spectrophotometric distance bin
over 4,800 square degrees. The errorbars are poissonian 1$\sigma$ errors
and the curve is the expected $d^2$ distribution, normalized at 16~pc.}
\label{distribution_function}
\end{figure}

where $\Phi$ is the luminosity function and $\sigma$ is the intrinsic rms
scatter in the spectral type-luminosity relation. The scatter in the $M_{\rm J}$ versus 
spectral type relation is $\sigma \sim 0.27$ mag (Fig. \ref{Mj_SpT}), which 
corresponds to a 7$\%$ overestimate of the stellar density. 

The mean surface density of our sample, $0.38 \pm 0.10$~objects per 100 
square degrees out to 22~pc, corresponds to an uncorrected luminosity 
function of 
$\overline{\Phi}_{\rm J}=(1.76 \pm 0.46) . 10^{-3}$~dwarfs~mag$^{-1}$~pc$^{-3}$. 
After correcting
for the Malmquist bias, this becomes 
$\overline{\Phi}_{\rm J~cor}=(1.64 \pm 0.46).10^{-3}$~dwarfs~mag$^{-1}$~pc$^{-3}$, averaged
over $11.1 \leq M_{\rm J} \leq 13.1$. Our value is in good agreement within error bars with
the \citet{cruz07} measurement of
$\overline{\Phi}_{\rm J}=(0.95 \pm 0.30).10^{-3}$~dwarfs~mag$^{-1}$~pc$^{-3}$, 
averaged over $11.25 \leq M_{\rm J} \leq 13.25$ (see their Table~11 and Fig.~17)
for M8--L4 ultracool dwarfs at high Galactic latitude ($|b| > 10\degr$).
This clearly demonstrates that the space density of ultracool dwarfs (M8--L3.5)
in the solar neighbourhood does not depend on Galactic latitude, as expected.
\section{Summary}
We discovered 20 nearby ultracool dwarfs M8--L5.5 with one L3.5 at only $\sim$10~pc
and reidentified six
known ones at low Galactic latitude. 
Four of them are low-pm ultracool dwarfs, giving
a significant fraction of $\sim 15\%$ of our sample. 
These ultracool dwarfs provide us additional
targets for detailed studies of the basic physical properties of stars at the bottom of the main
sequence.
\section*{Acknowledgments}
The DENIS project has been partly funded by the SCIENCE and the HCM plans of
the European Commission under grants CT920791 and CT940627.
It is supported by INSU, MEN and CNRS in France, by the State of Baden-W\"urttemberg 
in Germany, by DGICYT in Spain, by CNR in Italy, by FFwFBWF in Austria, by FAPESP in Brazil,
by OTKA grants F-4239 and F-013990 in Hungary, and by the ESO C\&EE grant A-04-046.
Jean Claude Renault from IAP was the Project manager.  Observations were  
carried out thanks to the contribution of numerous students and young 
scientists from all involved institutes, under the supervision of  P. Fouqu\'e,  
survey astronomer resident in Chile.  

The Digitized Sky Surveys were produced at the Space 
Telescope Science Institute under U.S. Government 
grant NAG W-2166. 
The images of these surveys are based on photographic 
data obtained using the Oschin Schmidt Telescope 
on Palomar Mountain and the UK Schmidt Telescope. 
The plates were processed into the present compressed 
digital form with the permission of these institutions. 
The National Geographic Society - Palomar Observatory 
Sky Atlas (POSS-I) was made by the California Institute 
of Technology with grants from the National Geographic Society. 
The Second Palomar Observatory Sky Survey (POSS-II) was 
made by the California Institute of Technology with 
funds from the National Science Foundation, the 
National Geographic Society, the Sloan Foundation, 
the Samuel Oschin Foundation, and the Eastman Kodak 
Corporation. The Oschin Schmidt Telescope is operated 
by the California Institute of Technology and Palomar 
Observatory. The UK Schmidt Telescope was operated by 
the Royal Observatory Edinburgh, with funding from the 
UK Science and Engineering Research Council (later the 
UK Particle Physics and Astronomy Research Council), 
until 1988 June, and thereafter by the Anglo-Australian 
Observatory. The blue plates of the southern Sky Atlas 
and its Equatorial Extension (together known as the SERC-J), 
as well as the Equatorial Red (ER), and the Second Epoch 
[red] Survey (SES) were all taken with the UK Schmidt.

This publication makes use of data products from
the Two Micron All Sky Survey, which is a joint project 
of the University of Massachusetts and Infrared
Processing and Analysis Center/California Institute of 
Technology, funded by the National Aeronautics
and Space Administration and the National Science 
Foundation; the NASA/IPAC Infrared Science Archive, which
is operated by the Jet Propulsion Laboratory/California 
Institute of Technology, under contract with the 
National Aeronautics and Space Administration.

This research has made use of the ALADIN, SIMBAD and VIZIER databases, 
operated at CDS, Strasbourg, France. 

We thank the referee for useful comments that significantly improved our paper.

\begin{landscape}
%\label{SpT}
\begin{center}
\tablehead{
\multicolumn{17}{p{22.5cm}}{{\bf Table 6.} 26 nearby L, and late-M dwarfs}\\
\hline\hline
\noalign{\smallskip}
 DENIS name  &  PC3 &  err  & TiO5  & err    & VOa & err   & Sp.T   & err & Sp.T    & err & Sp.T   & Sp.T    & err & Sp.T    & Sp.T      & Refs. \\
             &      & (PC3) &       & (TiO5) &     & (VOa) & (PC3)  &     & (TiO5)  &     & (VOa)  & (index) &     & (temp.) & (adopted) &       \\
 (1) & (2) & (3) & (4) & (5) & (6) & (7) & (8) & (9) & (10) & (11) & (12) & (13) & (14) & (15) & (16) & (17) \\\hline\noalign{\smallskip}} % end of tablehead
\begin{supertabular}{lrlllllllllllllll}
D* 0615$-$0100 &   3.73 &  0.48 & 0.871 & 0.211 & 2.05 & 0.43 & L2.3 & 0.4 & L1.2 & 1.2 & L2.3 & L2.0 & 1.0 & L2.5 & L2.5 &   \\
D* 0630$-$1840 &   2.63 &  0.20 & 0.336 & 0.030 & 2.35 & 0.30 & L0.1 & 1.4 & M8.1 & 0.2 & L0.2 & M9.0 & 1.0 & M8.0 & M8.5 &   \\
D* 0644$-$2841 &   2.38 &  0.29 & 0.315 & 0.067 & 1.92 & 0.63 & M9.7 & 0.8 & M8.0 & 0.4 & L3.2 & M9.0 & 0.5 & M9.5 & M9.5 & 1 \\
D* 0652$-$2534 &   2.28 &  0.22 & 0.768 & 0.040 & 2.27 & 0.24 & M9.5 & 0.6 & L0.6 & 0.2 & L0.7 & L0.0 & 0.5 & M9.5 & L0.0 &   \\
D* 0716$-$0630 &   3.01 &  0.32 & 0.590 & 0.166 & 2.45 & 0.65 & L1.7 & 1.5 & M9.6 & 0.9 & M9.5 & L0.5 & 1.0 & L1.0 & L1.0 &   \\ 
D* 0751$-$2530 &   3.79 &  0.40 & 0.985 & 0.116 & 2.29 & 0.33 & L2.4 & 0.3 & L1.8 & 0.7 & L0.6 & L2.0 & 0.5 & L1.0 & L1.5 &   \\ 
D* 0805$-$3158 &   1.95 &  0.16 & 0.180 & 0.114 & 2.19 & 0.24 & M8.5 & 0.7 & M7.2 & 0.6 & M6.8 & M8.0 & 0.5 & M8.0 & M8.0 &   \\ 
D* 0812$-$2444 &   3.11 &  0.38 & 1.086 & 0.151 & 2.28 & 0.49 & L1.8 & 0.3 & L2.4 & 0.9 & L0.7 & L2.0 & 0.5 & L1.0 & L1.5 &   \\ 
D* 0823$-$4912 &   3.52 &  0.68 & 0.885 & 0.077 & 2.07 & 0.57 & L2.1 & 0.6 & L1.2 & 0.4 & L2.2 & L1.5 & 0.5 & L1.0 & L1.5 &   \\ 
D* 0828$-$1309 &   2.54 &  0.32 & 0.750 & 0.201 & 2.16 & 0.44 & L0.0 & 1.5 & L0.5 & 1.1 & L1.5 & L0.5 & 1.5 & L1.0 & L1.0 & 2 \\
D* 1048$-$5254 &   2.91 &  0.49 & 0.874 & 0.242 & 2.77 & 0.77 & L1.6 & 1.8 & L1.2 & 1.4 & M7.3 & L1.5 & 1.5 & L1.0 & L1.5 &   \\ 
D* 1126$-$5003 &  11.03 &  8.17 & 0.772 & 0.395 & 1.46 & 0.98 & L5.9 & 4.3 & L0.6 & 2.2 & L6.5 & L6.0 & 4.5 & L5.0 & L5.5 & 3 \\  
D* 1157$-$4844 &   2.54 &  0.33 & 0.819 & 0.102 & 2.48 & 0.57 & L0.0 & 1.5 & L0.9 & 0.6 & M9.3 & L0.5 & 1.0 & L1.0 & L1.0 &   \\ 
D* 1159$-$5247 &   2.16 &  0.14 & 0.429 & 0.017 & 2.21 & 0.26 & M9.2 & 0.5 & M8.7 & 0.1 & M7.0 & M9.0 & 0.5 & M9.0 & M9.0 & 4 \\
D* 1232$-$6856 &   1.94 &  0.16 & 0.217 & 0.014 & 2.68 & 0.79 & M8.4 & 0.7 & M7.5 & 0.1 & L1.9 & M8.0 & 0.5 & M8.0 & M8.0 &   \\ 
D* 1253$-$5709 &   2.22 &  0.42 & 0.987 & 0.336 & 2.43 & 0.66 & M9.3 & 1.5 & L1.8 & 1.9 & M9.6 & L0.5 & 1.5 & L0.0 & L0.5 &   \\ 
D* 1347$-$7610 &   2.28 &  0.74 & 0.498 & 0.536 & 2.33 & 2.17 & M9.5 & 2.9 & M9.0 & 3.0 & L0.3 & M9.5 & 3.0 & L0.0 & L0.0 & 5 \\
D* 1454$-$6604 &   4.60 &  1.76 & 1.293 & 0.121 & 1.79 & 1.05 & L3.0 & 1.5 & L3.6 & 0.7 & L4.1 & L3.5 & 1.0 & L3.5 & L3.5 &   \\ 
D* 1519$-$7416 &   2.17 &  0.28 & 0.192 & 0.074 & 2.22 & 0.43 & M9.2 & 1.0 & M7.3 & 0.4 & M7.1 & M8.5 & 0.5 & M9.0 & M9.0 &   \\ 
D* 1520$-$4422 &   2.52 &  0.68 & 0.826 & 0.163 & 1.73 & 0.50 & L0.0 & 2.0 & L0.9 & 0.9 & L4.6 & L0.5 & 1.5 & L1.0 & L1.0 & 5 \\
D* 1705$-$5441 &   2.06 &  0.57 & 0.007 & 0.011 & 2.58 & 1.44 & M8.9 & 2.6 & M6.3 & 0.1 & L0.9 & M9.0 & 2.5 & M8.0 & M8.5 &   \\   
D* 1733$-$1654 &   2.62 &  0.40 & 0.725 & 0.081 & 2.17 & 0.48 & L0.1 & 1.6 & L0.3 & 0.5 & L1.5 & L0.0 & 1.0 & L1.0 & L1.0 &   \\ 
D* 1745$-$1640 &   3.23 &  0.73 & 0.636 & 0.157 & 1.95 & 0.71 & L1.9 & 1.9 & M9.8 & 0.9 & L3.0 & L1.0 & 1.5 & L2.0 & L1.5 &   \\ 
D* 1756$-$4518 &   1.97 &  0.20 & 0.310 & 0.177 & 2.38 & 0.36 & M8.5 & 0.8 & M8.0 & 1.0 & M8.8 & M8.5 & 1.0 & M9.0 & M9.0 &   \\    
D* 1756$-$4805 &   2.40 &  0.39 & 0.450 & 0.270 & 2.42 & 0.60 & M9.8 & 1.7 & M8.8 & 1.5 & M9.7 & M9.5 & 1.5 & L0.0 & L0.0 &   \\ 
D* 1909$-$1937 &   2.58 &  0.89 & 1.851 & 1.283 & 2.29 & 1.38 & L0.1 & 2.7 & L6.7 & 7.3 & L0.7 & L0.0 & 2.5 & L2.0 & L1.0 &   \\   
\noalign{\smallskip}\hline
\end{supertabular}
\end{center}

Abbreviations.---D*: DENIS\\
Col. (1): abbreviated DENIS name.
Cols. (2)--(7): spectroscopic indices and associated errors. PC3 defined \citet{martin99}; TiO5 defined
in \citet{reid95}; VOa defined in \citet{kirk99}.
Cols. (8)--(12): spectral types derived from the PC3, TiO5, and VOa index using the formula given
in \citet{martin99} and \citet{cruz02} and their associated errors (only for PC3 and TiO5).  
Cols. (13) \& (14): the mean spectral type derived based on PC3 and TiO5 and its associated errors
(see Section~6.2).
Col. (15): spectral type estimated from comparison with templates. 
Col. (16): the adopted spectral type (see Section~6.2).
Col. (17): references; 
(1): 2MASS J0644143$-$284141, \citet{cruz03}; 
(2): SSSPM J0829$-$1309, \citet{scholz02}; 
(3): 2MASS J11263991$-$5003550, \citet{folkes}; 
(4): 1RXS  J115928.5$-$524717, \citet{ham04};
(5): \citet{kendall}, for 2MASS J15200224$-$4422419 see also \citet{burl}. \\
\end{landscape}

\begin{table*}
\setcounter{table}{6}
 \caption{26 nearby L, and late-M dwarfs}
\label{dist}
 $$
 \begin{tabular}{lllrlrrrl}
   \hline 
   \hline
   \noalign{\smallskip}
 DENIS name  & Sp.T & $M_{\rm J}$ & d$_{\rm sp}$ & err & V$_{\rm t}$ & err & {\it EW} & Refs. \\ 
             &      &             &  (pc)        &(pc) &  km s$^{-1}$& km s$^{-1}$ &H$\alpha$(\AA) &    \\
 (1) & (2) & (3) & (4) & (5) & (6) & (7) & (8) & (9)  \\
   \hline
   \noalign{\smallskip}  
DENIS 0615$-$0100 & L2.5 & 12.74 & 16.0 & 3.3 &  18 &  5 &   $<$7.0 &   \\
DENIS 0630$-$1840 & M8.5 & 11.28 & 19.3 & 3.9 &  56 & 12 &  $<$18.0 &   \\
DENIS 0644$-$2841 & M9.5 & 11.65 & 26.8 & 5.6 &  44 & 13 &  $<$10.0 & 1 \\
DENIS 0652$-$2534 & L0.0 & 11.83 & 14.9 & 3.0 &  17 &  4 &   $<$1.0 &   \\
DENIS 0716$-$0630 & L1.0 & 12.19 & 22.0 & 4.6 &  13 &  3 &   $<$4.0 &   \\
DENIS 0751$-$2530 & L1.5 & 12.38 & 14.8 & 2.9 &  63 & 13 &   $<$5.0 &   \\
DENIS 0805$-$3158 & M8.0 & 11.10 & 26.4 & 5.4 &  32 &  7 &   $<$3.0 &   \\
DENIS 0812$-$2444 & L1.5 & 12.38 & 20.1 & 4.3 &  18 &  5 &   $<$5.0 &   \\
DENIS 0823$-$4912 & L1.5 & 12.38 & 17.4 & 3.6 &  11 &  3 &   $<$5.0 &   \\
DENIS 0828$-$1309 & L1.0 & 12.19 & 12.7 & 2.5 &  33 &  8 &   $<$4.0 & 2 \\
DENIS 1048$-$5254 & L1.5 & 12.38 & 21.0 & 4.6 &  18 &  5 &   $<$3.0 &   \\
DENIS 1126$-$5003 & L5.5 & 13.83 & 10.6 & 2.2 &  82 & 17 &  $<$12.0 & 3 \\
DENIS 1157$-$4844 & L1.0 & 12.19 & 23.3 & 4.9 &   6 &  3 &   $<$2.0 &   \\
DENIS 1159$-$5247 & M9.0 & 11.47 &  9.6 & 1.9 &  49 & 10 &   15.0   & 4 \\
DENIS 1232$-$6856 & M8.0 & 11.10 & 18.0 & 3.5 &  18 &  4 &   $<$6.0 &   \\
DENIS 1253$-$5709 & L0.5 & 12.01 & 19.4 & 4.0 & 150 & 31 &   $<$7.0 &   \\
DENIS 1347$-$7610 & L0.0 & 11.83 & 24.9 & 5.2 &  23 &  7 &  $<$10.0 & 5 \\
DENIS 1454$-$6604 & L3.5 & 13.10 & 10.5 & 2.1 &  32 &  7 &   $<$5.0 &   \\
DENIS 1519$-$7416 & M9.0 & 11.47 & 25.4 & 5.2 &  40 & 10 &   $<$2.0 &   \\
DENIS 1520$-$4422 & L1.0 & 12.19 & 16.2 & 3.2 &  56 & 12 &   $<$4.0 & 5 \\
DENIS 1705$-$5441 & M8.5 & 11.28 & 27.2 & 5.7 &  10 &  3 &   $<$6.0 &   \\
DENIS 1733$-$1654 & L1.0 & 12.19 & 19.2 & 4.0 &   9 &  4 &   $<$2.0 &   \\
DENIS 1745$-$1640 & L1.5 & 12.38 & 18.7 & 4.1 &  14 &  5 &   $<$9.0 &   \\
DENIS 1756$-$4518 & M9.0 & 11.47 & 14.8 & 3.0 &  14 &  3 &   $<$2.0 &   \\
DENIS 1756$-$4805 & L0.0 & 11.83 & 20.0 & 4.3 &   9 &  3 &  $<$13.0 &   \\
DENIS 1909$-$1937 & L1.0 & 12.19 & 26.9 & 6.0 &  20 & 10 &   $<$5.0 &   \\
    \noalign{\smallskip}
    \hline
   \end{tabular}
 $$
\begin{list}{}{}
\item[]
Col. (1): abbreviated DENIS name.
Col. (2): spectral type determined in this paper.
Col. (3): $J$-band absolute magnitude computed from spectral type using the calibration in this paper.
Cols. (4) \& (5): spectrophotometric distance and its associated error.
Cols. (6) \& (7): tangential velocity and its associated error.
Col. (8): H$\alpha$ equivalent widths (\AA).
Col. (9): references; 
(1): 2MASS J0644143$-$284141, \citet{cruz03}; 
(2): SSSPM J0829$-$1309, \citet{scholz02}; 
(3): 2MASS J11263991$-$5003550, \citet{folkes}; 
(4): 1RXS  J115928.5$-$524717, \citet{ham04};
(5): \citet{kendall}, for 2MASS J15200224$-$4422419 see also \citet{burl}. \\
\end{list}
\end{table*}
%

%
%\label{lastpage}
\end{document}